\documentclass{aa}  
\usepackage{graphicx}
\usepackage{txfonts}
\usepackage{natbib}
\bibpunct{(}{)}{;}{a}{}{,}
\begin{document}
   \title{Evidence for dust accumulation just outside the orbit of Venus}

   \subtitle{}

   \author{Ch. Leinert\inst{1}
           \and
           B. Moster\inst{1}
          }

   \offprints{leinert@mpia.de}

   \institute{Max-Planck-Institut f\"ur Astronomie,
              K\"onigstuhl 17, D-69117 Heidelberg, Germany
             }

   \date{Received xxx xx, 2006; accepted xxx xx, 2006}

% \abstract{}{}{}{}{} 
% 5 {} token are mandatory
 
  \abstract
   {}
   {To contribute to the knowledge of dynamics of interplanetary dust
    by searching for structures in the spatial distribution of 
    interplanetary dust near the orbit of Venus.}
   {We study the radial gradient of zodiacal light brightness,
   as observed by the zodiacal light photometer on board the
   Helios space probes on several orbits from
   1975 to 1979. The cleanest data result from Helios B
   (= Helios 2) launched in January 1976.}
   {With respect to the general increase of zodiacal light brightness
    towards the Sun, the data show an excess brightness of a few percent
    for positions of the Helios space probe just outside the orbit of Venus.}
   {We consider this as evidence for a dust ring associated with
    the orbit of Venus, somewhat similar to that found earlier along
    the Earth's orbit.}

   \titlerunning{Dust ring outside Venus' orbit}
   \maketitle  
%
%________________________________________________________________

\section{Introduction}

The orbital evolution of interplanetary dust with its typical sizes
of 1 $\mu$m to 100 $\mu$m is dominated by the Poynting-Robertson-effect.
This braking effect resulting from
the aberration of solar radiation drives the
grains into the Sun in astronomically short times, about 10$^5$ years
for a 10 $\mu$m particle originating at 3 AU distance from the Sun
in the asteroid belt \cite{wyatt50}. Unless interplanetary dust is
a transitory phenomenon of our times, this requires continuous
replenishment. Most of the source region for interplanetary dust particles
lies in the asteroid belt and beyond \cite{dermott2001}. 
While spiralling inwards, 
these interplanetary dust particles will cross the orbits of the inner 
planets where gravitational perturbations of these planets may produce
detectable signatures in the overall spatial distribution of
interplanetary dust. Indeed, a circumsolar dust ring along the Earth's
orbit producing an excess of about 10\%  in zodiacal light brightness
was detected in the infrared zodiacal light measurements
of IRAS and COBE \cite{reach2002}. Numerical simulations 
\cite{dermott1994,dermott2001} have explained the main features
by resonances experienced by particles in low-inclination orbits,
probably originating from sources in the asteroid belt. 
It is expected that
similar effects might be seen with the next inner planet, Venus.
We therefore decided to search the zodiacal light data of the
Helios space probes \cite{leinert1981a}, 
which crossed the orbit of Venus, for
a brightness excess of
similar size which could show the presence
of a density enhancement, or ``dust ring'' along the orbit of Venus.
To our knowledge this is the only data set which adresses this
issue.
Such a study is appropriate now, whith the study of a Solar Probe mission 
into the outer
solar atmosphere signaling renewed interest in the inner solar system.
Also, the necessary knowledge
of basics and details of the operation of the Helios space
probes and their zodiacal light photometer instrument is still
available.
%
%________________________________________________________________

\section{Instrument}

Helios is a spacecraft spinning with one revolution per s with the 
spin axis perpendicular to its orbital plane which is the ecliptic.
The zodiacal light is measured by three photometers mounted rigidly into
the spacecraft. One --  the 90$^{\circ}$-photometer -- looks 
parallel to the
spin axis towards the ecliptic pole with a circular filed-of-view of 
 3$^{\circ}$ diameter. The other two -- the 15$^{\circ}$-photometer and
the 30$^{\circ}$-photometer -- are mounted at angles such that they scan 
small circles of ecliptic latitudes $\beta$ = 16$^{\circ}$ and
$\beta$ = 31$^{\circ}$ with instantaneous square fields-of-view of
1$^{\circ}$ x 1$^{\circ}$ and 2$^{\circ}$ x 2$^{\circ}$. For Helios A 
the three photometers were looking south, for Helios B
they were looking north of the ecliptic \cite{leinert1975}.

Observations were performed in the visual in the U, B, and V bands.
As detector a photon counting photomultiplier of type EMR 541N was used.
An integration of the 90$^{\circ}$-photometer lasted 126 s. The signal
of the 360$^{\circ}$ scans performed by the 15$^{\circ}$- and 
30$^{\circ}$-photometers was integrated into 32 sectors of width 
5.6$^{\circ}$, 11.2$^{\circ}$ and 22.5$^{\circ}$ over 513 revolutions,
resulting in integration times per sector of 8 s, 16 s or 32 s respectively.
The signals were strong,
with typically 2x10$^4$ to 4x10$^5$
counted photoelectrons contributing to each of the
individual measurements.
For a concise description of the instrument see Leinert et al. (1975).

One of the strong points of the zodiacal light instrument is its stability.
Apart from a slow trend related to the orbital temperature variation
inside the spacecraft, the stability measured over a month 
on an internal  calibration lamp \cite{leinert1981b} or on the sky
\cite{richter1982} was better than 0.5\% with respect to the smooth
average trend, and the long-term stability over the whole 11 year lifetime of
Helios A was about 0.1\% per year \cite{leinert1989}. It is this
stability which makes it worthwhile at all to search for effects
of density structure in the interplanetary dust cloud near the orbit of
Venus. By analogy to the Earth's resonant ring such structures -- if they
exist -- are expected to result
in brightness structures of only a few percent.

\section{Orbit}

Helios A and Helios B were launched on December 10, 1974 and January 16,
1976 into elliptical heliocentric orbits with perihelia at 0.31~AU and
0.29~AU, respectively. The aphelia, according to the launch dates, were at
heliocentric ecliptic longitudes $\lambda$
of 87$^{\circ}$ and 114$^{\circ}$.
While the Helios orbits closely follow the ecliptic plane, the orbit
of Venus is inclined by 3.4$^{\circ}$ {with a line of nodes at
$\Omega$ = 75.5$^{\circ}$. At this heliocentric longitude - 
and again 180$^{\circ}$ later - Helios will cross the orbital plane of Venus.
However, these positions are not relevant for our study.
What counts, is the times when Helios is close to the {\em orbit}
of Venus. In practice, we identify as points of closest approach those
positions where the Helios spacecraft are crossing the orbit of Venus}.
 This leads to
four crossings with the Venus orbit, two for each spacecraft, {one on the
inbound, one on the outbound part of the Helios orbit. 
Given the inclination of Venus' orbit},
its distance z$_V$ from the ecliptic
- north counted positive - {will usually be non-negligible, and it will
have} different values for the different crossings.
The relevant
geometrical parameters are summarised in Table~\ref{tab_orbit_crossing}.
A northern position of the orbit of
Venus is favourable when searching a brightness 
enhancement with the north-looking
zodiacal light photometers on Helios B, unfavourable for the
south-looking photometers on Helios A.

\begin{table}
\caption{Gemometry of the crossings of Venus' orbit \label{tab_orbit_crossing}}
\begin{tabular}{lrrrc}
\hline
         & R~(AU)  & $\lambda$~($^{\circ}$) & z$_V$(AU) 
                                                     & conditions for\\
         &         &                        &        &observing excess\\
\hline
Helios A & \\
\hspace*{1.5em}inbound & 0.717 & 126.4 & 0.033 & unfavourable\\
\hspace*{1.5em}outbound & 0.733 & 30.6 & -0.036 & favourable\\
Helios B & \\
\hspace*{1.5em}inbound & 0.717 & 160.2 & 0.042 & favourable\\
\hspace*{1.5em}outbound & 0.721 & 67.9 & -0.006 & neutral\\
\hline
\end{tabular}
\end{table}

The orbital period of the Helios spacecraft of 190~d and 185~d
are shorter than the 225~d period of Venus, therefore the position of Venus  
in its orbit changes by about -60$^{\circ}$ from one orbit crossing
of Helios to the same crossing an Helios orbit later.
Indeed, for one of the crossings (Helios A inbound on orbit 3 on May 8, 1976)
Venus was by chance in the field-of-view of 
sector 12 of the 15$^{\circ}$- photometer.
 Since the inbound 
and outbound crossings also differ by $\sim$100$^{\circ}$ in $\lambda$,
our measurements below, averaged over several orbits, will not refer
to a special position with respect to Venus but to average conditions
along the Venus orbit.

Not all of the Helios orbits could be used for our study. Because of
increasing solar activity, only orbits before the second half of 1978
were sufficiently free of solar flare and interplanetary plasma
induced effects \cite{leinert1989}. Parts of the first orbit were
affected by attitude maneuvers which abruptly changed viewing
direction and therefore observed brightness.

\section{Geometry}

The geometry for the 90$^{\circ}$-photometers is simple. While Helios is 
crossing a density enhancement, the zodiacal light photometer will locally
observe the effects of any excess dust in the column along the line of
sight towards the ecliptic pole. For the 15$^{\circ}$- and 
30$^{\circ}$ - photometers with their oblique lines-of-sight the geometry
is more involved. The best chances to detect an enhancement will be for those
sectors where the line-of-sight is oriented almost tangentially to
the orbit of Venus. These are in the numbering of the instrument
the sectors 12 and 21, the centers of which are at
$\lambda - \lambda_{\odot}$ = $\pm$ 84.4$^{\circ}$ (sector 12) and
$\lambda - \lambda_{\odot}$ = $\mp$ 84.4$^{\circ}$ (sector 21). Here,
the upper sign is valid for Helios A, the lower one for Helios B
with its flipped spin and orientation. The zodiacal light brightness in
these directions is 2-2.5 times higher than at the north ecliptic
pole, but the longer oblique light path through a density enhancement
compensates for this dilution by
contributions from distant dust.

\section{Data}

We concentrate our search on data from Helios B. Helios A has the disadvantage
that its 90$^{\circ}$ photometer is pointing into the outer regions of
the Magallanic cloud and that - due to to the failed deployment of one of 
its two antennas - the spacecraft and with it the field-of-view, e.g. of the
90$^{\circ}$ photometer is wobbling at each revolution with an amplitude
of about 0.5$^{\circ}$. Therefore no standard reduction was attempted
on these data. They are quality data only in the innermost
solar system part of the Helios orbit where the zodiacal light is up
to 17 times brighter than when close to the Earth. Also, 
the wobble mentioned above increases
the uncertainty of corrections for individual stars in the 
15$^{\circ}$- and 30$^{\circ}$- photometers.

Thus we have a distinct north-south gradient in the systematic reliability of the data:
those obtained with the 90$^{\circ}$ photometer on Helios B are clearly the
best, and the main conclusions should be based on them. The other photometers on Helios B
are more subject to uncertainties in corrections of individual stars and
of galactic background. We included those data sets which were apparently 
taken under favourable conditions concerning the contributions of
background and individual stars. The consistency of these results 
with the data of the
90$^{\circ}$ photometer strongly adds to our findings.
Data from Helios A were less reliable at
the level of percent accuracy for the reasons given above.
We show those two examples which were still taken under relatively good conditions.
They appear broadly consistent with the results of Helios B. This is
a weaker but nevertheless favourable contribution to the overall picture.

\section{Data reduction}

Calibration and data reduction are described in Leinert et al. (1981a, 1981b)
and are not repeated here. 
We take the data as they were reduced by the Helios team during the time of
the mission and stored at the data center NSSDC. The quantity we use 
is the calibrated zodiacal light brightness given in S10 units
obtained from the total
signal by subtracting dark current (negligible), light scattered by the
electrons of the interplanetary medium (small), the contributions 
of individual stars (mostly small, sometimes dominating) and from
integrated starlight (never small, often major).
In the V band one S10 unit - 
the equivalent of one solar type star of magnitude 10
per square degree - corresponds to 
1.18\,10$^{-8}$ Wm$^{-2}$sr$^{-1}$$\mu$m$^{-1}$ or 1200 Jy/sr.

\begin{figure}[t!]  % ----------------------------------- Figure 1
\centering
\includegraphics[width=88mm,angle=0]{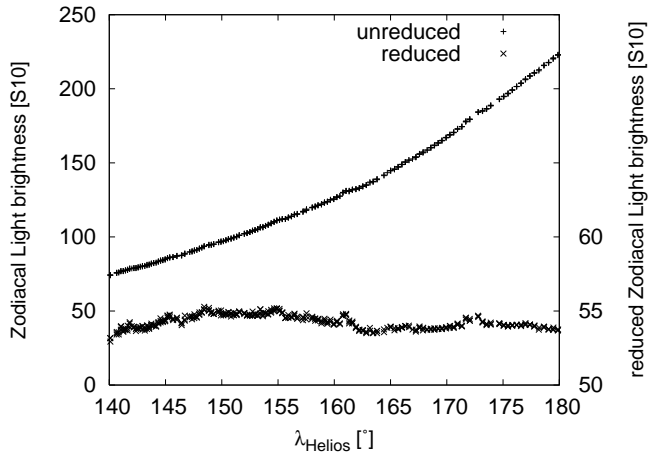}
\caption{Zodiacal light brightness (upper curve) and reduced
brightness (lower curve) as seen with the 90$^{\circ}$ photometer
on Helios B in V during the inbound crossing of Venus' orbit in 
March 1977. The reduced brightness was obtained from the 
original results by removing the gradient (see text). The 
heliocentric longitude $\lambda_{\rm Helios}$ gives the position of 
Helios along its orbit. Note the expanded scale with suppressed zero,
which was used for the reduced brightness to show the remaining
structure.
}
\label{fig_multiply_R}
\end{figure}

Helios data reduction is non-recursive. For the reduction, the data were 
used as observed without removing or
smoothing or interpolation. The reduction is based on literature
values for star brightnesses and stellar background brightnesses and
was done with calibrations and transformations to the colour system
of the zodiacal light photometer as determined in the laboratory.
Incorrect input brightnesses from the catalogues used
or errors in spacecraft attitude
will show as disturbances in the data. The disadvantages of this
method are obvious. The advantage is that the result is 
transparently close
to what the raw observations show with little improving art applied. 
This facilitates using the data again, now three decades after the initial,
authorised
reduction.

\begin{figure}[t!]   % ------------------------------------ Figure 2
\centering
\includegraphics[width=88mm,angle=0]{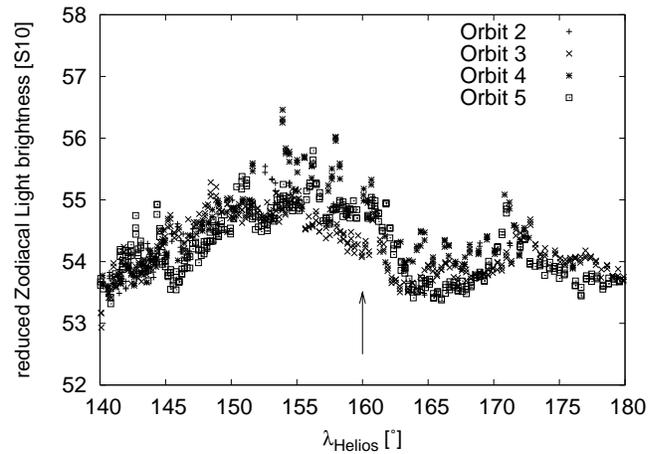}
\caption{Repeatability of the observed structure in reduced
zodiacal light brightness from orbit to orbit of the Helios B
space probe. The figure shows The V band observations at the 
north ecliptic pole. The arrow indicates the position of Helios B at 
the inbound crossing of the orbit of Venus.
}
\label{fig_orbit_repeat}
\end{figure}

\begin{figure}[b!]  % ------------------------------- Figure 3
\centering
\includegraphics[width=88mm,angle=0]{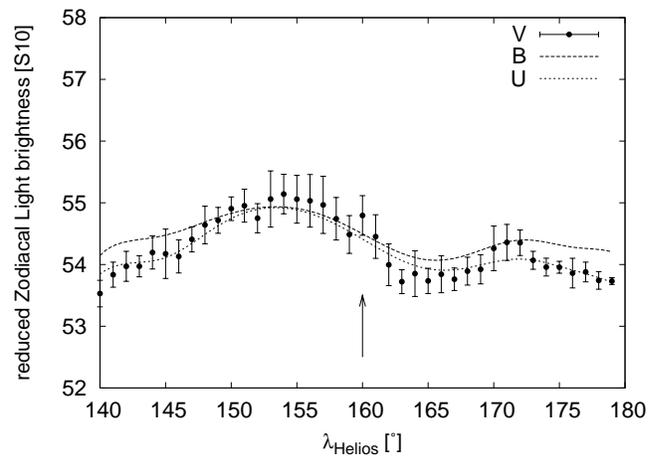}
\caption{Same as Fig. 2 but with the data averaged in 1$^{\circ}$
intervals in heliocentric longitude, and with the observations in U and B
superimposed. The error bars, shown for V only, represent the rms scatter of
the typically 10-20 measurements per bin. This is a quite conservative
error, since for purely statistical
fluctuations, the errors of the mean in each bin would be smaller by a factor
of 3-4. Here, as in the following figures, the curves
showing the averages of the U and B measurements have been shifted 
by one to a few S10 for ease of comparison, if necessary. The arrow marks the
position of Helios B at the crossing.
}
\label{fig_excess_solar_colour}
\end{figure}

\section{Results}

The most obvious feature of Helios zodiacal light measurements
is the strong brightness increase towards the Sun, approximately 
$\sim$~R$^{-2.3}$, where R is the Sun-Helios distance. Therefore,
the data were multiplied by a factor R$^{\alpha}$  with exponent
$\alpha$ close to 2.3 in order to
produce a time series with little variation in which 
it would be easier to search for
possible systematic variations near the orbit of Venus
(Fig.~\ref{fig_multiply_R}).

\begin{figure}[t!]  % ----------------------------------- Figure 4
\centering
\includegraphics[width=88mm,angle=0]{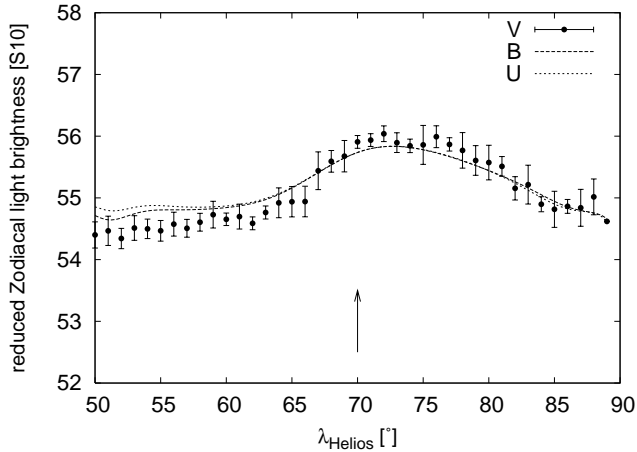}
\caption{Enhancement in reduced zodiacal light brightness
near the outbound crossing of Venus' orbit as seen in the observations
of Helios B towards the north ecliptic pole. Again, measurements from
orbits 2-5 have been averaged. The arrow marks the position of Helios B 
at the crossing.
}
\label{fig_90deg_outbound}
\end{figure}

In the following we always refer to this {\em reduced brightness} which
has no precisely defined meaning but approximately gives the
zodiacal light brightness at 1 AU.
The multiplication by the smooth  large-sale factor R$^{\alpha}$  
does not introduce artefacts
of measurable size. As noted above, the best data are available for 
{\em Helios B}.

We start with the results obtained with its 90$^{\circ}$ photometer,
since these -- measuring locally the
contribution from a column above the spacecraft
-- provide the most direct probe for enhancements
in spatial density of interplanetary dust. These data 
do show a small brightness enhancement just outside the orbit of Venus,
as suggested in  Fig.~\ref{fig_multiply_R}. In the following we
check the credibility of this effect.

If this brightness enhancement is due to a structure in the interplanetary
dust cloud, one would expect it to repeat during each orbit at the
same orbital position. Figure~\ref{fig_orbit_repeat} shows for the
observations towards the north ecliptic pole and
for the inbound crossing that this is the
case. 

One would also expect that the enhancement has approximately
solar colour, as is true for the zodiacal light. In other words
this  means that
in U, B, and V the brightness excess - measured in S10 units should have the
same size. This condition is also fulfilled as demonstrated 
in Fig.~\ref{fig_excess_solar_colour}.

Finally, on the outbound crossing of Venus' orbit, -- as expected -- again an enhancement
occurs just outside the orbit of Venus (Fig.~\ref{fig_90deg_outbound}).

\begin{figure}[t!] % -------------------------------- Figure 5
\centering
\includegraphics[width=88mm,angle=0]{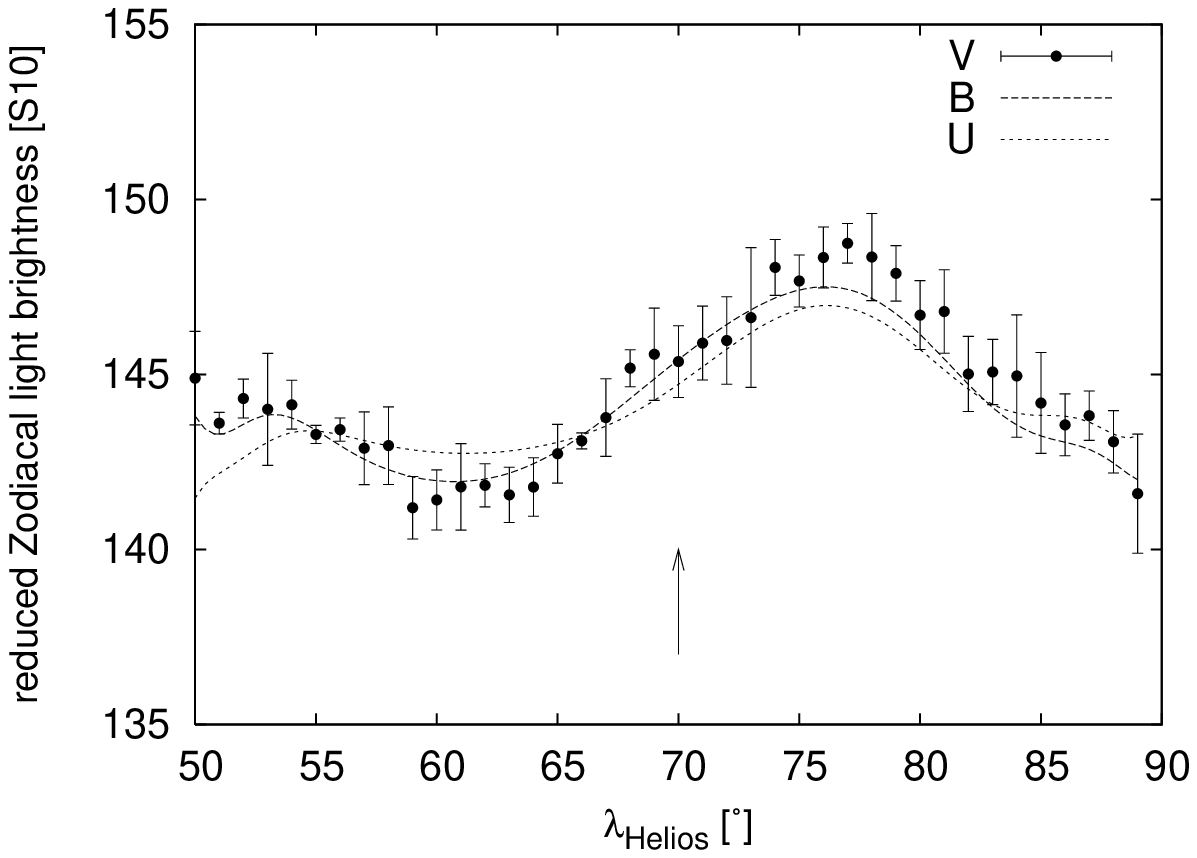}
\rule{0pt}{4mm}
\includegraphics[width=88mm,angle=0]{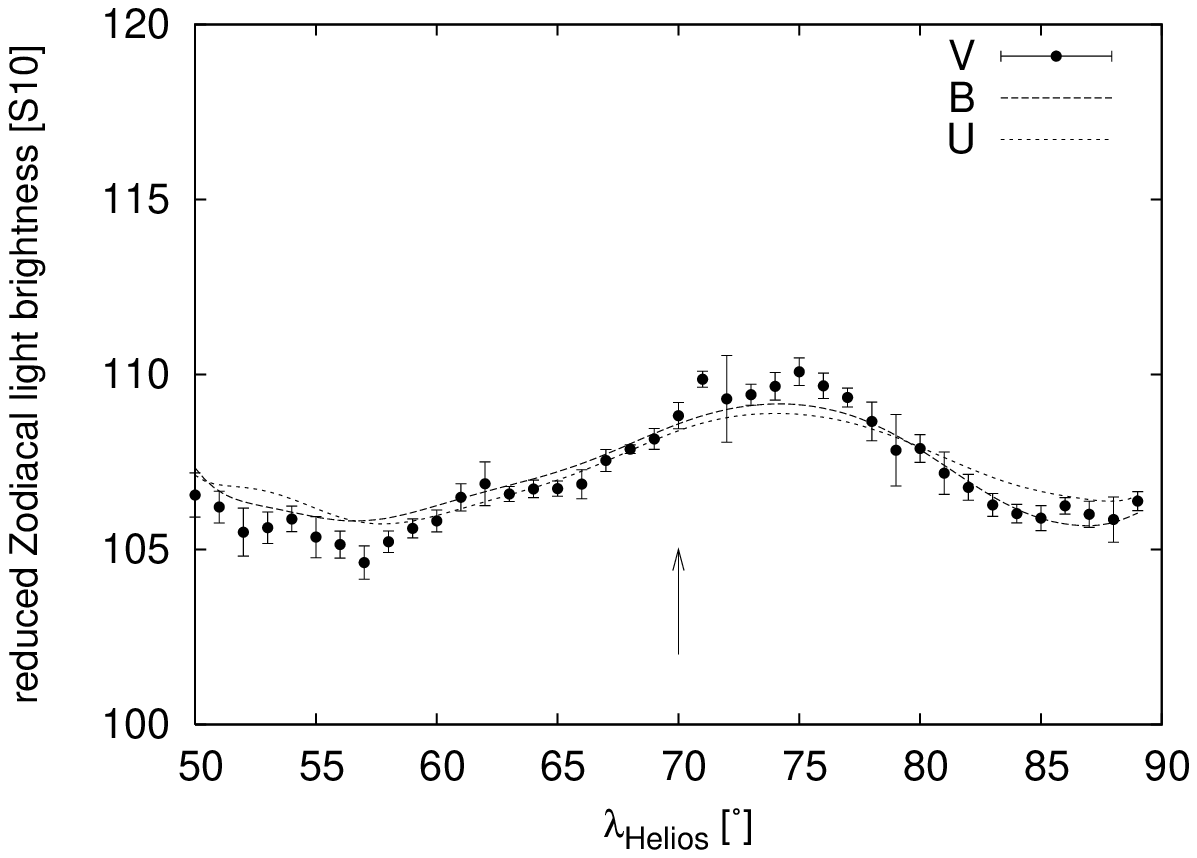}
\caption{Brightness enhancement seen with the other photometers of 
Helios B on the outbound crossing of Venus' orbit. Upper panel: 
sector 12 of the photometer looking at ecliptic latitude $\beta$
$\approx$ 15$^{\circ}$. Bottom: sector 12 of the photometer looking 
at ecliptic latitude $\beta$ $\approx$ 30$^{\circ}$. Same
presentation as in Figs.~\ref{fig_90deg_outbound} and
~\ref{fig_excess_solar_colour}, but the average is over orbits
1-5. The arrow marks the position of Helios B at the crossing.
}
\label{fig_B15_30_outbound_S12}
\end{figure}

\begin{figure}[h!] % --------------------------------- Figure 6
\centering
\rule{0pt}{4mm}
 \includegraphics[width=88mm,angle=0]{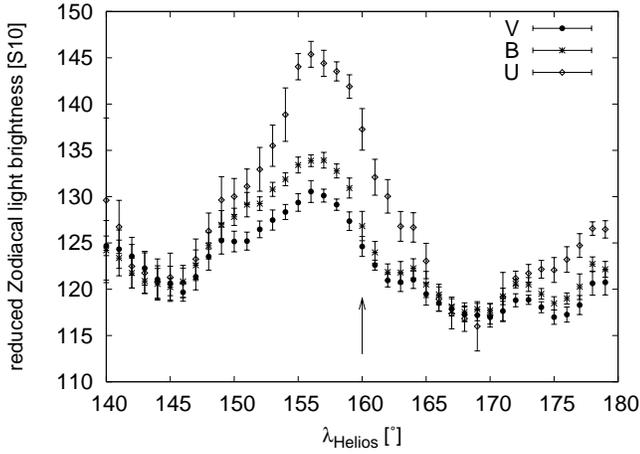}
\caption{Example of a false enhancement, observed by Helios B in sector 21
of the photometer looking at $\approx$ 15$^{\circ}$ during the inbound 
crossing. The excess, rather blue, can be traced to the B1V star HR 1074.
}
\label{fig_stellar_disturbance_B15}
\end{figure}

It is virtually impossible that these effects are produced by stars passing 
through the field of view of the photometer. This field is always pointed 
towards the north ecliptic pole, with a small offset of 0.5$^{\circ}$
which precesses along a circle around the 
ecliptic pole within one orbital period
of Helios B, or with a motion of 0.09$^{\circ}$ over the duration
of the observed brightness enhancement. The excess
brightness of about 2 S10 in the 3$^{\circ}$ diameter field-of-view
could by produced by a star of magnitude V=7.1 with solar colours fully entering and
leaving again the field-of-view at exactly the right times.
No star this bright is present anywhere near the edge of the field-of-view, 
and 
even if it
were, it could not move in and out again with such a small,
one-directional displacement.

The 15$^{\circ}$- and 30$^{\circ}$ - photometers of Helios B are more prone to spurious
brightness changes because of the varying stellar and diffuse background.
However two of the eight combinations (two latitudes, two sectors, two
crossings) both have low background and no bright stars in the field-of-view
and therefore are expected to give reliable results.
These, both referring to the outbound crossing of Venus' orbit, are shown
in Fig.~\ref{fig_B15_30_outbound_S12}.

%The other ``disturbed'' scans are summarised
%in Fig.~\ref{fig_B_bad_scans}. It is no surprise that in the presence
%of brighter stars and in regions of a galactic background in excess of
%50-100 S10 such disturbances will occur. In some (most?) of them
%one can still imagine that signals like shown above are present
%beneath the disturbing fluctuations.

Figure~\ref{fig_stellar_disturbance_B15} shows a counterexample.
Here the brightness excess observed close to the orbit of Venus 
-- as can be concluded from its colour -- is mostly an artefact due to incomplete
subtraction of stellar contributions. In this particular case it
looks like the stellar excess signal due to the B1V star HR~1074
is superimposed to a $\approx$ 4 S10 underlying zodiacal light
brightening similar to that seen in Fig~\ref{fig_B15_30_outbound_S12}.
But a decomposition of the signal into these two contributions might 
be stressing the data. (With sufficient optimism one can guess the presence
of a zodiacal light excess brightness signal also in the other disturbed
data sets).

\begin{figure}[t!]   % ----------------- Figure 7
\centering
\includegraphics[width=88mm,angle=0]{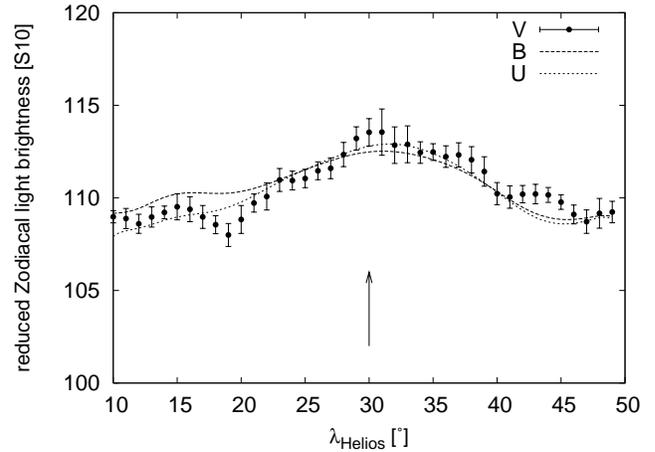}
\caption{Brightness enhancement seen with Helios A in sector 12 of the
  photometer looking at $\approx$ 30$^{\circ}$ during the outbound crossing
of Venus' orbit. 
}
\label{fig_helios_a_outbound}
\end{figure}

\begin{figure}[h!]   % ------------------- Figure 8
\centering
\includegraphics[width=88mm,angle=0]{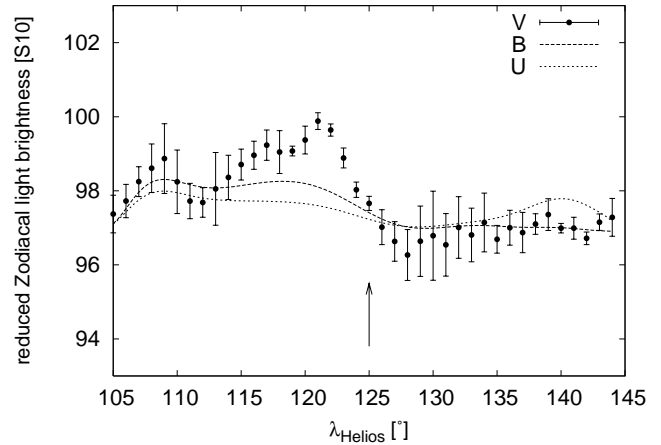}
\caption{Brightness variation seen with Helios A in sector 12 of the
  photometer looking at $\approx$ 30$^{\circ}$ during the inbound crossing
of Venus' orbit. There is some discrepancy between the measurements in
the different colours, but at comparatively low level.
}
\label{fig_helios_a_inbound}
\end{figure}

With {\em Helios A} the conditions for detecting small brightness 
excesses are less favourable than for Helios B for the instrumental reasons
mentioned above. In addition, there is a tendency for higher backgrounds and
brighter stars in its south-looking photometers. The least interference by
bright stars is for the outbound crossing shown in
Fig.~\ref{fig_helios_a_outbound}. There is also a measurement during an
inbound crossing of Venus' orbit with low background (Fig.~\ref{fig_helios_a_inbound}),
but here the results for V are somewhat discrepant from the results in B and U.
One could conclude that no zodiacal light enhancement in excess
of 2 S10 is evident here.

In Table \ref{table_results_summary} we summarise the data sets with
observed enhancements
in zodiacal light brightness near the orbit of Venus and indicate why
for other data sets reliable data could not be obtained.

\begin{table}
\caption{Excesses in reduced zodiacal light brightness
observed near the orbit of Venus}
\label{table_results_summary}
\begin{tabular}{llrrrr}
Helios & viewing  &
\multicolumn{4}{c}{inbound $\leftarrow$ crossing $\rightarrow$ outbound} \\
 & direction& [S10]& \% & [S10]& \%\\
\hline
 B & 90$^{\circ}$ & 1.2$\pm$0.3 & 2 & 1.2$\pm$0.3 & 2\\
  & 31$^{\circ}$ sec. 12 & \multicolumn{2}{l}{high background} & 3.6$\pm$0.8   & 3.5\\
  & 31$^{\circ}$ sec. 21 & \multicolumn{2}{l}{$\alpha$ Per at edge of FoV}&
                             \multicolumn{2}{l}{high background}\\
  & 16$^{\circ}$ sec. 12 & 
    \multicolumn{2}{l}{bright star $\epsilon$ Oph}& 5.8$\pm$1.0   & 4\\
  & 16$^{\circ}$ sec. 21 & \multicolumn{2}{l}{bright star at edge FoV}     &
     \multicolumn{2}{l}{bright star $\Theta$ Peg} \\
\hline
 A   & -16$^{\circ}$ sec. 12 & \multicolumn{2}{l}{Mira, $\eta$, 
    $\Theta$ Cet in FoV}   &\multicolumn{2}{l}{high background}  \\
  & -16$^{\circ}$ sec. 21 & \multicolumn{2}{l}{V and B inconsistent} 
     &\multicolumn{2}{l}{Procyon in FoV} \\
  & -31$^{\circ}$ sec. 12 &  \multicolumn{2}{l}{V and B inconsistent}   &  3.7$\pm$0.9    & 3.5\\
  & -31$^{\circ}$ sec. 21 & \multicolumn{2}{l}{high background}  &  
        \multicolumn{2}{l}{high background} \\
\hline
%%\multicolumn{6}{l}{stars: 5\% of star signal should be $<$ 1/4 signal - absolu%%te}\\
%%\multicolumn{6}{l}{3\% of background should be $<$ 1/4 signal - relative}\\
\end{tabular}
\end{table}

\section{Discussion}

The results of the previous section, presented in Figs. 3-5, 7 and 8,
 show a brightness enhancement
by a few percent immediately outside the orbit of Venus. The interpretation 
depends on whether or not one accepts the reasons given in favour of an
origin of this enhancement by structures in the spatial distribution
of interplanetary dust.

\subsection{Upper limits to an excess in dust density near
the orbit of Venus}

Upper limits are the result of our study if the relation of the brightness excess to
underlying dust structures is not accepted, but the brightness excess
is attributed to some other non-specified effect. Any brightness fluctuations 
due to dust structures then have to be significantly less than the observed
effect. Therefore a safe upper bound to possible enhancements of dust density
near the orbit of Venus can be derived from the observed brightness
increases. These are
2\% for a viewing direction at $\beta$ = +90$^{\circ}$, 4\%
for a viewing direction at $\beta$ = +16$^{\circ}$ and longitude
$\lambda$ - $\lambda _{\odot}$ = 84$^{\circ}$, and 3.5\%
for a viewing direction at $\beta$ = $\pm$31$^{\circ}$ and longitude
$\lambda$ - $\lambda _{\odot}$ = $\mp$84$^{\circ}$. Using a result from 
the following subsection, this would correspond to less than 10\% overdensity 
in any  dust structure
extending over about 0.05 AU in direction perpendicular to the
ecliptic plane.

We consider this as a useful, not easy to
perform observational achievement,
and we leave it to others to determine how discriminating these limits may
be for descriptions of orbital evolution of interplanetary dust
particles.

Indeed our original intent was to derive such an upper limit
for the smoothness of the interplanetary dust cloud near the
orbit of Venus. However, the data suggested the
existence of a non-random feature.

\subsection{Simple model -- dust ring outside the orbit of Venus}

A dust ring is the result of our study if the relation of the brightness increase to
underlying dust structures is accepted.
To this end we have to verify whether the different detections of
a small brightness increase could be caused by one single
spatial structure.

For this test a very simple model will suffice: a ring just ouside the
orbit of Venus with radial extent $\Delta$R and thickness $\Delta$z,
symmetrical to the orbital plane of Venus, and in which the spatial
density of interplanetary dust is increased by x\% with respect to the
smooth overall radial distribution.

{The fact that the symmetry plane of interplanetary dust deviates
from the ecliptic plane does not present a problem to this
simple model but rather supports it. 
For the inner solar system, the parameters of this
symmetry plane were determined as i = 3.0 $\pm$ 0.3$^{\circ}$,
$\Omega$ =  87 $\pm$ 4$^{\circ}$ \cite{leinert1980}, almost
coincident with the orbital plane of Venus. This gives some
physical justification to assuming a ring model with symmetry to
Venus' orbital plane.
}

\subsubsection{Predictions from the simple model}

The radial extent $\Delta$R follows from the duration of the event
and the velocity of Helios along its orbit.
The full duration was about 11$^{\circ}$ in $\lambda_{Helios}$ corresponding to
$\Delta$R $\sim$ 0.08 AU. The FWHM extent measures about
8$^{\circ}$ in $\lambda_{Helios}$ corresponding to  $\Delta$R $\sim$ 0.06 AU.
This appears to be the same for all detections.

Values for $\Delta$z and x are most directly connected to the
observations with the 90$^{\circ}$-photometer, which shows an
excess in reduced zodiacal light brightness of 1.2 S10 or 2\%.
The local scattering power of interplanetary dust at 1 AU and
scattering angle 90$^{\circ}$ was determined
using quite general arguments from the brightness gradient in the
ecliptic  by Dumont (1973). Dumont\&Levasseur-Regourd (1978) 
noted that therefore
for observations to the north ecliptic pole at 1 AU, 1.8 S10 are contributed 
by scattering along the first 0.01 AU of of the line-of-sight.
If the radial distribution of interplanetary dust follows a power law,
which is true to good approximation in the inner solar system, similarity 
relations hold and the same value of 1.8 S10 per now 0.01$\cdot$$R_{Helios}$
AU also
applies for the observed reduced brightnesses. We cannot determine the
quantities $\Delta$z and x separately from our observations, but
just {\em assuming} x = 10\%, a ring thickness of $\Delta$z = 0.048 AU
would be necessary to produce the observed excess brightness. The 
inbound crossing of Helios B occurs 0.04 AU south of Venus' orbit and 
therefore will observe the full 1.2 S10 of this excess when looking
to the north ecliptic pole. 

For the lines-of-sight of the 15$^{\circ}$- and 30$^{\circ}$ - photometer,
with $\beta$ = 16.2$^{\circ}$ and 31.3$^{\circ}$, the part of the
line-of-sight crossing the ring of enhanced dust density has the
length $\Delta$s = $\Delta$z/sin($\beta$) = 0.17 AU, respectively 0.09 AU.
Since the first one or two tenths of an AU along the lines-of-sight the
scattering angle is almost constant, $\approx$ 90$^{\circ}$, 
the expected excess brightness in these photometers
is larger than that observed with the 90$^{\circ}$ - photometer by
the factor 1/sin($\beta$), resulting in predicted brightness enhancements
of 4.3 S10 in the
15$^{\circ}$- and 2.3 S10 in the 30$^{\circ}$ - photometer. 
%%This is close 
%%to the observed values (see Table~\ref{tab_observed_excesses}).

Smaller values are predicted for crossings where the line-of-sight
does not penetrate the full thickness of the assumed ring because
the orbit of Helios cuts the ring or passes it on the wrong side.
Thus for the outbound crossing of Helios B the brightness excess should be
smaller by about a factor of two, and for the inbound crossing of Helios A,
virtually no enhancement should be seen. This last prediction could be best
checked with the data of the 90$^{\circ}$ - photometer of Helios A,
which unfortunately are not useable for the technical reasons given above.

\subsubsection{Comparison with the results}

The observations during the outbound orbit crossing of Helios B provide
a remarkable data set. Not only shows the 1.2 S10 excess seen in the 
90$^{\circ}$ - photometer solar colour, but the excesses observed with 
the independent 15$^{\circ}$-  and 30$^{\circ}$ - photometers also
show excesses, and not only in the same colour but also in the same
length along the Helios orbit and almost in the brightness ratios expected for
a circumorbital ring of increased spatial content.

The inbound crossing at first glance looks consistent, showing the 
same brightness excess. But at least for the simple ring model the observed 
excesses are too small, they should show at about twice the strength at
the outbound crossing.

According to the orbital geometry given in Table~\ref{tab_orbit_crossing},
the south-looking photometers on Helios A should see no or almost no 
brightness excess near the orbit of Venus on the inbound crossing but should
experience the full signal on the outbound part. Unfortunately, the data
of the 90$^{\circ}$ - photometer, which would give the most direct result -
are not useable here. From Fig.~\ref{fig_helios_a_outbound} we see an excess 
of similar size as for Helios B on the outbound crossing. The excess here
seems
to be located rather on than outside the orbit of Venus, but this crossing is
also at a somewhat larger heliocentric distance than the others
(Table~\ref{tab_orbit_crossing}). An inbound crossing
of Helios A is shown in Fig.~\ref{fig_helios_a_inbound}. Here the data in V
show a structure not present in the bands B and U and therefore do not really
qualify as reliable measurements. Hence we can only state that
the data appear not inconsistent with the predicted absence of a clear
brightness excess for this crossing.

We do not have as convincing proof as we would like. However the
evidence is best where the data are cleanest, namely for the pole-viewing 
90$^{\circ}$ - photometer of Helios B. In addition,
in view of the smallness of the effect
and of the known difficulty to correct for the stellar contributions in the
15$^{\circ}$-  and 30$^{\circ}$ - photometers our interpretation
of the data is that the observed excesses are a real feature of
the zodiacal light and that they provide reasonably strong evidence for
the existence of a ring of enhanced dust content just
outside the orbit of Venus.

\subsection{Future research}

It would be natural to supplement these observations by specific 
dynamical studies of the orbital evolution of interplanetary dust
particles near the orbit of Venus. However, this is beyond the scope of this
paper. We limit ourselves to provide the observational input,
to which we feel qualified,  and leave the theoretical
part to the specialists.

A search for a dust enhancement near the orbit of Mercury was not
attempted. The low mass of this planet and the strong solar irradiation,
resulting in a much faster Poynting-Robertson drift would act to
keep any Mercury-induced enhancement of dust density very small.

It would be an interesting task for a future inner solar system space probe
with imaging capabilities to try to obtain a  picture of
the suggested dust ring which is more complete 
than was possible with the local
measurements of the Helios space probes at a total of four orbital crossroads.

\section{Summary and conclusion}

We searched the data of the zodiacal light photometer on the
Helios A and and in particular on the Helios B space probe
 for a brightness excess close to the
orbit of Venus with the following results:

\begin{itemize}

\item [--] we found a small excess, 2\% towards the north ecliptic pole,
which is repeatable from orbit to orbit.
For the outbound crossing of Helios B, the excess fulfills all of the following
criteria which should hold if it was due to a local enhancement in
interplanetary dust: solar colour, same orbital extent for the observed
ecliptic latitudes of $\beta$ = 16$^{\circ}$, 31$^{\circ}$,
and   90$^{\circ}$, and excess brightness varying with 1/sin($\beta$).

\item [--]interpreted by a dust ring just outside the orbit of Venus and
symmetrical to its orbital plane,
the radial width of this ring would be 0.06 AU FWHM. Its height would be
0.048 AU for an assumed overdensity of interplanetary dust in this ring
of 10\%.

\item [--]this very suggestive evidence is somewhat weakened by the fact that
the excesses on the outbound crossing of Helios B have the same strength
as for the inbound crossing, contrary to the prediction of this very simple
ring model. Obviously this simple ring model cannot be the
complete story.

\item[--] the data for Helios A are less reliable, but 
 appear consistent with the effects produced by such a dust ring.

\end{itemize}

All in all, the evidence is sufficiently convincing - maybe not
beyond any doubt - to suggest the existence of a ring of
enhanced dust content just outside the orbit of Venus. It will be
worthwhile to study the implications of this finding for the
dynamics of interplanetary dust particles and to compare
to the already known dust ring along the Earth's orbit.

\begin{acknowledgements}
We thank very much the many colleagues from the Helios project, in particular
Eckhart Pitz and Hartmut Link, the co-PIs of the zodiacal light photometer instrument. By building the instrument and spacecraft,
by performing operations, data transfer and data reduction these many, working
tediously over a long time, provided the
basis for this work. We gratefully acknowledge the efforts of the
National Space Science Data Center NSSDC, which kept the data from being lost,
and we thank very much Bernie Jackson from UCSD who brought these data
into easily useable format. In particular we want to mention H. Czech from
Dornier System who designed the high voltage supplies of the zodiacal light
photometer to be by a substantial factor more stable and more repeatable
than specified -- without this gift we could not even have thought of
starting this study.
\end{acknowledgements}

\end{document}